\documentclass[%
 reprint,
 amsmath,amssymb,
 aps,
]{revtex4-2}

\usepackage{graphicx}
\usepackage{dcolumn}
\usepackage{bm}

\begin{document}

\preprint{APS/123-QED}

\title{Comment on ``Generalized James' Effective Hamiltonian Method"}

\author{W. Rosado}
\email{wilson.rosado@unisucre.edu.co}
\affiliation{%
 GIFTA, Departamento de Física, Universidad de Sucre,\\
 Cra 28 No 5-267 Puerta Roja, Sincelejo, Colombia}%


\author{Ivan Arraut}
\email{ivan.arraut@usj.edu.mo}
\affiliation{
University of Saint Joseph,\\
Estrada Marginal da Ilha Verde, 14-17, Macao, China
}%


\date{\today}

\begin{abstract}
In the paper carried out by Wenjun et al. \cite{Shao2017}, a generalization of the James effective dynamics theory based on a first version of the James method was presented. This however, is not a very rigorous way of deriving the effective third-order expansion for an interaction Hamiltonian with harmonic time-dependence. In fact, here we show that the third-order Hamiltonian obtained in \cite{Shao2017} is not Hermitian for general situations when we consider time-dependence. Its non-Hermitian nature arises from the foundation of the theory itself. In this comment paper, the most general expression of the effective Hamiltonian expanded up to third order is obtained. Our derived effective Hamiltonian is Hermitian even in situations where we have time-dependence.
\end{abstract}

\maketitle


\section{\label{sec:level1} Introduction}

The effective Hamiltonian method of James and Jerke \cite{James2007} has been widely applied to physical systems related to radiation-matter interactions in the large-detuning regime \cite{Cho2008,Liu2016,Cao2018,Ivanov2020,Wang2022,RamosPrieto2020,Su2021,Yuan2018,Li2023,Rosado2015,Bin2022}.
While there are other methods for finding
effective dynamics with applications in quantum information theory and quantum optics
 \cite{Ren2019,Lee2018,Gardiner2004,Klimov2002,Reiter2012}, James’ method
stands out for providing a compact expression for obtaining an effective Hamiltonian equivalent to the second-order interaction Hamiltonian of the system. However, this expression is only valid if the atom-field coupling is
sufficiently small such that we can neglect higher-order terms inside the Dyson expansion of the time evolution
operator. It was precisely this weak-coupling condition that motivated Wenjun et al. \cite{Shao2017} to propose a generalization of James’ effective Hamiltonian method and then offered an expression of the effective Hamiltonian expanded up to the third order for dealing with problems in the strong and ultra-strong atom-field coupling regime, as it is required by the physical processes governed by counter-rotating terms in Rabi’s quantum model inside the large detuning regime \cite{Ma2015,Garziano2016}. The same approximation is valid in situations where the Jaynes-Cummings model (weak coupling) is applied \cite{Zhao2017}. However, the method used to extract the desired effective Hamiltonian was based on the first version of the James method (see appendix in \cite{James2000}), which has been proved to be not so rigorous. Consequently, the Hermiticity of the dynamics obtained by this method cannot be guaranteed in general, particularly for the cases involving time-dependent effective dynamics. The most recent versions of James' effective dynamics method are based on a time-averaged dynamics of the system \cite{James2007,James2010}, which is similar to a low-pass filter dividing the Hilbert space into low and high frequencies. This improved method is telling us that we must be careful with the Hermiticity of the effective Hamiltonian in order to describe unitary dynamics for a closed quantum system. This care is not considered in \cite{Shao2017}. Therefore, the effective Hamiltonian to the third order obtained in \cite{Shao2017} is not Hermitian as we will demonstrate in a moment. In \cite{Shao2017}, it was proved that the Hamiltonian is Hermitian for a particular case where the effective Hamiltonian is independent of time, but not in general. \par The effective dynamics with harmonic time-dependence can be of great interest in quantum state engineering \cite{Rosado2015,Liu2016,Bin2022,Prado2006}. This is the case because these oscillations allow new transformations or very specific interaction representations of the effective Hamiltonian, which are necessary for the preparation and protection of the target quantum state. This in addition demonstrates the necessity to provide a correct expression for the effective third-order Hamiltonian which guarantees its Hermiticity. Therefore, it is extremely important to make the corresponding corrections.

\section{\label{sec:level1} The Hermiticity Problem in the Generalized James' Effective Hamiltonian Method}

We briefly review the method presented in \cite{Shao2017}, to expose the difficulties that it presents in the Hermiticity of the proposed effective Hamiltonian. For the dynamics analyzed inside the scenario of the interaction picture, we start with a Hamiltonian having with the following structure:

\begin{equation}
\hat{H}_{I}\left(t\right)=\sum_{m}\left[\hat{h}_{m}^{\dagger}e^{i\omega_{m}t}+\hat{h}_{m}e^{-i\omega_{m}t}\right]
\label{eq:interaction_hamiltonian},
\end{equation}
In \cite{Shao2017}, it is proposed an iterative solution of the Schr\"odinger equation from the solution
\begin{equation}
\left|\varphi\left(t\right)\right\rangle =\left|\varphi\left(0\right)\right\rangle +\frac{1}{i\hbar}\int_{0}^{t}\hat{H}_{I}\left(t^{\prime}\right)\left|\varphi\left(t^{\prime}\right)\right\rangle dt^{\prime},
\end{equation}
and after iterating \textit{n} times, removing highly oscillating terms $\hat{H}_{I}\left(t\right)\left|\varphi\left(0\right)\right\rangle$, and using the Markov approximation, an effective Hamiltonian of order \textit{n} is obtained, i.e.
\begin{eqnarray}   \hat{H}_{eff}^{\left(n\right)}\left(t\right)=&&\left(\frac{1}{i\hbar}\right)^{n-1}\hat{H}_{I}\left(t\right)\int_{0}^{t}\hat{H}_{I}\left(t_{1}\right)\int_{0}^{t_{1}}\hat{H}_{I}\left(t_{2}\right)\nonumber\\
    &&\times
    \cdots\times\int_{0}^{t_{n}-2}\hat{H}_{I}\left(t_{n-1}\right)dt_{n-1}\cdots dt_{2}dt_{1}.
\end{eqnarray}
Thus, the effective third-order Hamiltonian is given by
\begin{equation}
  \hat{H}_{eff}^{\left(3\right)}=\left(\frac{1}{i\hbar}\right)^{2}\hat{H}_{I}\left(t\right)\int_{0}^{t}\hat{H}_{I}\left(t_{1}\right)\int_{0}^{t_{1}}\hat{H}_{I}\left(t_{2}\right)dt_{2}dt_{1} 
  \label{eq:Heff_third_order_wenjun}
\end{equation}
If the interaction Hamiltonian is of the form defined in equation (\ref{eq:interaction_hamiltonian}), then the effective Hamiltonian in equation (\ref{eq:Heff_third_order_wenjun}) becomes
\begin{widetext}
\begin{eqnarray}
\hat{H}_{eff}^{\left(3\right)}\left(t\right)=&&\frac{1}{\hslash^{2}}\sum_{n,m,k}\frac{1}{\omega_{k}\left(\omega_{k}-\omega_{m}\right)}\left[\hat{h}_{n}\hat{h}_{m}\hat{h}_{k}^{\dagger}e^{-i\left(\omega_{n}+\omega_{m}-\omega_{k}\right)t}+\hat{h}_{n}^{\dagger}\hat{h}_{m}\hat{h}_{k}^{\dagger}e^{i\left(\omega_{n}-\omega_{m}+\omega_{k}\right)t}\right]\nonumber\\
&&+\frac{1}{\hslash^{2}}\sum_{n,m,k}\frac{1}{\omega_{k}\left(\omega_{k}-\omega_{m}\right)}\left[\hat{h}_{n}\hat{h}_{m}^{\dagger}\hat{h}_{k}e^{-i\left(\omega_{n}-\omega_{m}+\omega_{k}\right)t}+\hat{h}_{n}^{\dagger}\hat{h}_{m}^{\dagger}\hat{h}_{k}e^{i\left(\omega_{n}+\omega_{m}-\omega_{k}\right)t}\right]\nonumber\\
&&+\frac{1}{\hslash^{2}}\sum_{n,m,k}\frac{1}{\omega_{k}\left(\omega_{m}+\omega_{k}\right)}\left[\hat{h}_{n}^{\dagger}\hat{h}_{m}\hat{h}_{k}e^{i\left(\omega_{n}-\omega_{m}-\omega_{k}\right)t}+\hat{h}_{n}\hat{h}_{m}^{\dagger}\hat{h}_{k}^{\dagger}e^{-i\left(\omega_{n}-\omega_{m}-\omega_{k}\right)t}\right]. 
 \label{eq:Heff_third_order_harmonic_wenjun}
\end{eqnarray}
\end{widetext}
The Hamiltonian defined in the equation (\ref{eq:Heff_third_order_harmonic_wenjun}) has been tested by the authors of \cite{Shao2017} in recent studies, showing then the simultaneous excitation of two atoms with a single photon \cite{Garziano2016} and the coupling of three photons in the Rabi quantum model within the large-detuning regime \cite{Ma2015}. In both results, the author in \cite{Shao2017}, succeeded in describing exactly the same effective dynamics. However, the Hamiltonian used for analyzing the system is not Hermitian, at least not up to the order of the approximation used for the analysis. Its non-Hermiticity is derived from the Hamiltonian in equation (\ref{eq:Heff_third_order_wenjun}). To understand this, let us consider a system where the third-order processes are relevant in comparison with those of lower order. In this case, the dynamics of the system would be completely determined by the effective Hamiltonian given in equation (\ref{eq:Heff_third_order_wenjun}) and then the time evolution of the system can be determined by means of the operator
\begin{equation}
    \hat{U}\left(t_{0}+dt,t_{0}\right)\approx1-\frac{i}{\hbar}\hat{H}_{eff}^{\left(3\right)}dt,
    \label{eq:Time_Evol_Operator}
\end{equation}
where $\hat{U}\left(t_{0}+dt,t_{0}\right)$, is the time evolution operator expanded in power series up to the first order of $dt$. To check the unitarity of equation. (\ref{eq:Time_Evol_Operator}), we make the product
\begin{equation}
    \hat{U}\hat{U}^{\dagger}\approx\hat{1}+\frac{i}{\hbar}\left(\hat{H}_{eff}^{\dagger\left(3\right)}-\hat{H}_{eff}^{\left(3\right)}\right)dt.
\end{equation}
If we want a unitary time evolution for the system, then the condition $\hat{H}_{eff}^{\dagger\left(3\right)}=\hat{H}_{eff}^{\left(3\right)}$ has to be satisfied. It is interesting to notice that in order to guarantee this condition, the following result must be true
\begin{equation}
\hat{H}_{I}\left(t_{2}\right)\hat{H}_{I}\left(t_{1}\right)=\hat{H}_{I}\left(t_{1}\right)\hat{H}_{I}\left(t_{2}\right).
\end{equation}
However, the interaction Hamiltonian given in equation (\ref{eq:interaction_hamiltonian}), does not commute with itself at different times. Therefore, the Hermiticity of the effective third-order Hamiltonian in equation (\ref{eq:Heff_third_order_wenjun}) cannot be guaranteed and this issue is translated to the calculations of the effective Hamiltonian given in equation 
(\ref{eq:Heff_third_order_harmonic_wenjun}). In \cite{Shao2017}, the Hermiticity of equation (\ref{eq:Heff_third_order_harmonic_wenjun}) is proved. However in order to do so, the author assumes the particular case where the condition $\omega_{k}+\omega_{m}-\omega_{n}=0$ or $-\omega_{k}-\omega_{m}+\omega_{n}=0$ is valid, eliminating in this way all temporal dependence in equation (\ref{eq:Heff_third_order_harmonic_wenjun}). In other words, the previously mentioned conditions are only valid when we wish to obtain an effective interaction which is independent of time. Outside this condition, the Hamiltonian of equation (\ref{eq:Heff_third_order_harmonic_wenjun}), is no longer Hermitian. As it has been previously mentioned, in some situations, especially in the case of quantum state engineering, the construction of an effective time-dependent interaction allows rotations or transformations that provide specific interactions for the desired goal, and for that purpose, the Hamiltonian in equation (\ref{eq:Heff_third_order_harmonic_wenjun}) is not appropriate because it would generate inaccurate results. \\\\
Initially, D.F.V. James used an iterative method to solve the Schrödinger equation and obtained the effective Hamiltonian up to second order \cite{James2000}. However, this approach raises serious questions because truncating the Dyson series to obtain the desired order of effective interaction results in non-unitarity \cite{Brouder2008}, as we have already demonstrated. Owing to the aforementioned concerns and considering the growing interest of other authors in applying the James technique in \cite{James2000}, by then James and Jerke in \cite{James2007} and Gamel and James in \cite{James2010}, conducted a more rigorous investigation into the applicability of the Hamiltonian obtained through the iterative method. As a result, the effective dynamics of the time average was obtained, which corrected the issues of Hermiticity, as we will show in the following section. 

\section{Extension to third order James' method: time-averaged dynamics}
The latest versions of D.F.V James’ method in separate collaborations with Jerke and Gamel are based on time-averaged dynamic evolution. In \cite{James2007}, James and Jerke establish the theoretical foundations of the effective dynamics and later in an improved version, Gamel and James derived an evolution equation in the form of Lindblad’s open system dynamics that takes into account the effects of time averaging the system’s observables \cite{James2010}. Next, we briefly present James’ time-averaged dynamics to obtain the expression for the third-order effective Hamiltonian and how this Hamiltonian is verified by the latest version of Gamel and James.
\subsection{\label{sec:level2}Time-averaged dynamics by James and Jerke}
James' theory of effective Hamiltonians \cite{James2007} consists in analyzing the dynamic of the system by evaluating the time average of any quantum operator $\hat{\mathcal{O}}\left(t\right)$
\begin{equation}
    \overline{\hat{\mathcal{O}}\left(t\right)}=\intop_{-\infty}^{\infty}f\left(t-t^{\prime}\right)\hat{\mathcal{O}}(t^{\prime})dt^{\prime},
\end{equation}
where $f\left(t\right)\epsilon\mathbb{R}$ and acts as a low-pass filter, removing high-frequency terms from the average. Using the condition $\frac{\partial\overline{O(t)}}{\partial t}=\frac{\overline{\partial O(t)}}{\partial t}$, the time-averaged Schr\"odinger equation in the interaction picture is as follows
\begin{equation}
    i\hbar\frac{\partial}{\partial t}\overline{\hat{U}(t,t_{0})}=\hat{H}_{eff}\left(t\right)\overline{\hat{U}(t,t_{0})},
\end{equation}
where it is not difficult to notice that,
\begin{equation}
    \hat{\mathcal{H}}_{eff}\left(t\right)=\left(\overline{\hat{H}_{I}\hat{U}(t,t_{0})}\right)\left(\overline{\hat{U}(t,t_{0})}\right)^{-1}.
    \label{eq:Heff_non_hermitian}
\end{equation}
Since $\left(\overline{\hat{U}(t,t_{0})}\right)^{-1}$ is not unitary, then the Hamiltonian defined in equation (\ref{eq:Heff_non_hermitian}) is not Hermitian. Therefore, the effective Hamiltonian must be
\begin{equation}
    \hat{H}_{eff}=\frac{1}{2}\left\{ \mathcal{\hat{H}}_{eff}+\mathcal{\hat{H}}_{eff}^{\dagger}\right\}.
     \label{eq:Heff_hermitian_James}
\end{equation}
By substituting in equation (\ref{eq:Heff_hermitian_James}) the power series expansion of the time evolution operator, the effective Hamiltonian can be determined up to the desired order, thus for example, by expanding up to the third order, we have
\begin{equation}                    
\hat{H}_{eff}=\overline{\hat{H}_{I}}+\hat{H}_{eff}^{\left(2\right)}+\hat{H}_{eff}^{\left(3\right)},
\end{equation}
where 
\begin{equation}
    \hat{H}_{eff}^{\left(2\right)}=\frac{1}{2}\left\{ \overline{\left[\hat{H}_{I},\hat{U}_{1}\right]}-\left[\overline{\hat{H}_{I}},\overline{\hat{U}_{1}}\right]\right\},
\end{equation}
and
\begin{eqnarray}
    \hat{H}_{eff}^{\left(3\right)}	=&&\frac{1}{2}\left\{ \overline{\hat{H}_{I}}\overline{\hat{U}_{2}^{\dagger}}-\overline{\hat{H}_{I}\hat{U}}_{1}\overline{\hat{U}_{1}}+\overline{\hat{H}_{I}\hat{U}_{2}}+\overline{\hat{U}}_{2}\overline{\hat{H}_{I}}\right.\nonumber\\
    &&\left.-\overline{\hat{U}}_{1}\overline{\hat{U}_{1}\hat{H}_{I}}+\overline{\hat{U}_{2}^{\dagger}\hat{H}_{I}}\right\} 
     \label{eq:Heff_order3_hermitian_James}
\end{eqnarray}
In these previous expressions, we define $\hat{U}_{1}$ and $\hat{U}_{2}$ as the first and the second-order terms of the series expansion of the operator $\hat{U}(t,t_{0})$, given by
\begin{equation}
    \hat{U}_{1}(t)=-\frac{i}{\hslash}\int_{0}^{t}\hat{H}_{I}(t^{\prime})dt^{\prime},
\end{equation}
and 
\begin{equation}
    \hat{U}_{2}(t)=-\frac{1}{\hslash^{2}}\int_{0}^{t}dt^{\prime}\hat{H}_{I}(t^{\prime})\int_{0}^{t^{\prime}}dt^{\prime\prime}\hat{H}_{I}(t^{\prime\prime}).
\end{equation}
For an interaction Hamiltonian in the form defined in equation (\ref{eq:interaction_hamiltonian}), we have $\overline{\hat{H}_{I}}=0$ in the large-detuning regime. Thus, the Hamiltonian given in equation (\ref{eq:Heff_order3_hermitian_James}), reduces to
\begin{equation}
    \hat{H}_{eff}^{\left(3\right)}=\frac{1}{2}\left\{ \overline{\hat{U}_{2}^{\dagger}\hat{H}_{I}}-\overline{\hat{H}_{I}\hat{U}}_{1}\overline{\hat{U}_{1}}+\overline{\hat{H}_{I}\hat{U}_{2}}-\overline{\hat{U}}_{1}\overline{\hat{U}_{1}\hat{H}_{I}}\right\},
    \label{eq:Heff_order3_HI=0}
\end{equation}
and in terms of the operators $\hat{h}_{m}$, the Hamiltonian in equation (\ref{eq:Heff_order3_HI=0}) becomes
\begin{widetext}
\begin{eqnarray}
    \hat{H}_{eff}^{\left(3\right)}=&&\frac{1}{2\hslash^{2}}\sum_{n,m,k}\left[\frac{1}{\omega_{n}\left(\omega_{n}+\omega_{m}\right)}+\frac{1}{\omega_{k}\left(\omega_{k}-\omega_{m}\right)}\right]\left(\hat{h}_{n}^{\dagger}\hat{h}_{m}^{\dagger}\hat{h}_{k}e^{i\left[\omega_{n}+\left(\omega_{m}-\omega_{k}\right)\right]t}+\hat{h}_{n}\hat{h}_{m}\hat{h}_{k}^{\dagger}e^{-i\left[\omega_{n}+\left(\omega_{m}-\omega_{k}\right)\right]t}\right)\nonumber\\
    &&+\frac{1}{2\hslash^{2}}\sum_{n,m,k}\left[\frac{1}{\omega_{k}\left(\omega_{m}+\omega_{k}\right)}+\frac{1}{\omega_{n}\left(\omega_{n}-\omega_{m}\right)}\right]\left(\hat{h}_{n}^{\dagger}\hat{h}_{m}\hat{h}_{k}e^{i\left[\omega_{n}-\left(\omega_{m}+\omega_{k}\right)\right]t}+\hat{h}_{n}\hat{h}_{m}^{\dagger}\hat{h}_{k}^{\dagger}e^{-i\left[\omega_{n}-\left(\omega_{m}+\omega_{k}\right)\right]t}\right)\nonumber\\
    &&+\frac{1}{2\hslash^{2}}\sum_{n,m,k}\left[\frac{1}{\omega_{n}\left(\omega_{n}-\omega_{m}\right)}+\frac{1}{\omega_{k}\left(\omega_{k}-\omega_{m}\right)}\right]\left(\hat{h}_{n}^{\dagger}\hat{h}_{m}\hat{h}_{k}^{\dagger}e^{i\left[\omega_{n}-\left(\omega_{m}-\omega_{k}\right)\right]t}+\hat{h}_{n}\hat{h}_{m}^{\dagger}\hat{h}_{k}e^{-i\left[\omega_{n}-\left(\omega_{m}-\omega_{k}\right)\right]t}\right).
        \label{eq:Heff_order3_harmonic_James}
\end{eqnarray}
\end{widetext}
This Hamiltonian is completely Hermitian and it is not necessary to impose that the algebraic sum of the frequencies is zero. Therefore, unlike the Hamiltonian given in equation (\ref{eq:Heff_third_order_harmonic_wenjun}), it can be used to determine the effective time-dependent dynamic, as we will discuss in section \ref{cap:4_secc}.
\subsection{\label{sec:level2} Time-averaged dynamics by Gamel and James} A more rigorous derivation of the time-averaged dynamics was presented in \cite{James2010}. In this new version, the effects of eliminating high-frequency terms caused by time-averaged dynamics were considered. When high-frequency terms in the Hamiltonian are removed, information about high-frequency processes is also being removed. This loss of information is reflected in the appearance of decoherence terms in the evolution equations for the average, and they are so important that including them produces corrections that can bring us closer to the case of exact evolution, particularly if more than one frequency is considered in the system. In the appendix of \cite{James2010}, Gamel and James derived the Lindblad term to the third order (see equation (A6)). For Hamiltonians with harmonic time dependence in the dispersive regime, namely, where $\overline{\hat{H}_{I}}=0$, the equation (A6) in \cite{James2010} is reduced to the following expression:
\begin{eqnarray}
    \mathcal{L}_{3}\left[\bar{\hat{\rho}}\right]=&&\overline{\hat{H}_{I}\hat{U}_{2}}\hat{\rho}+\overline{\hat{H}_{I}\hat{\rho}\hat{U}_{2}^{\dagger}}-\hat{\rho}\overline{\hat{U}_{2}^{\dagger}\hat{H}_{I}}-\overline{\hat{U}_{2}\hat{\rho}\hat{H}_{I}}\nonumber
    \\&&+\overline{\hat{H}_{I}\hat{U}_{1}\hat{\rho}\hat{U}_{1}^{\dagger}}-\overline{\hat{H}_{I}\hat{U}_{1}}\hat{\rho}\overline{\hat{U}_{1}^{\dagger}}-\overline{\hat{H}_{I}\overline{\hat{U}_{1}}\hat{\rho}\hat{U}_{1}^{\dagger}}\nonumber
    \\&&-\overline{\hat{U}_{1}\hat{\rho}\hat{U}_{1}^{\dagger}\hat{H}_{I}}+\overline{\hat{U}_{1}}\hat{\rho}\overline{\hat{U}_{1}^{\dagger}\hat{H}_{I}}+\overline{\hat{U}_{1}\hat{\rho}\overline{\hat{U}_{1}^{\dagger}}\hat{H}_{I}}\nonumber
    \\&&-\overline{\hat{H}_{I}\hat{U}}_{1}\overline{\hat{U}_{1}}\hat{\rho}-\overline{\hat{H}_{I}\hat{\rho}\overline{\hat{U}_{1}^{\dagger}}\hat{U}_{1}^{\dagger}}+\hat{\rho}\overline{\hat{U}^{\dagger}_{1}}\overline{\hat{U}_{1}^{\dagger}\hat{H}_{I}}\nonumber
    \\&&+\overline{\hat{U}_{1}\overline{\hat{U}_{1}}\hat{\rho}\hat{H}_{I}},
    \label{eq:Lindblad_third_order}
\end{eqnarray}
where $\hat{\rho}$ is the density matrix of the system. Introducing the operator $\hat{B}=\overline{\hat{H}_{I}\hat{U}_{2}}-\overline{\hat{H}_{I}\hat{U}}_{1}\overline{\hat{U}_{1}}$, the equation can be rewritten as follows
\begin{equation}
\mathcal{L}_{3}\left[\bar{\hat{\rho}}\right]=\left[\hat{H}_{eff}^{\left(3\right)},\bar{\hat{\rho}}\right]+\left\{ \frac{\hat{B}-\hat{B}^{\dagger}}{2},\bar{\hat{\rho}}\right\} +\mathcal{D}_{3}\left[\bar{\hat{\rho}}\right],
\label{eq:Lindblad_third_order_2}
\end{equation}
where 
\begin{equation}
    \hat{H}_{eff}^{\left(3\right)}=\frac{\hat{B}+\hat{B}^{\dagger}}{2},
     \label{eq:Hamiltonian_third_order_lindblad}
\end{equation}
and where the decoherence terms are defined as
\begin{eqnarray}
    \mathcal{D}_{3}\left[\bar{\hat{\rho}}\right]=&&\overline{\hat{H}_{I}\hat{\rho}\hat{U}_{2}^{\dagger}}-\overline{\hat{U}_{2}\hat{\rho}\hat{H}_{I}}+\overline{\hat{H}_{I}\hat{U}_{1}\hat{\rho}\hat{U}_{1}^{\dagger}}-\overline{\hat{H}_{I}\hat{U}_{1}}\hat{\rho}\overline{\hat{U}_{1}^{\dagger}}\nonumber
    \\&&-\overline{\hat{H}_{I}\overline{\hat{U}_{1}}\hat{\rho}\hat{U}_{1}^{\dagger}}-\overline{\hat{U}_{1}\hat{\rho}\hat{U}_{1}^{\dagger}\hat{H}_{I}}+\overline{\hat{U}_{1}}\hat{\rho}\overline{\hat{U}_{1}^{\dagger}\hat{H}_{I}}\nonumber
    \\&&+\overline{\hat{U}_{1}\hat{\rho}\overline{\hat{U}_{1}^{\dagger}}\hat{H}_{I}}-\overline{\hat{H}_{I}\hat{\rho}\overline{\hat{U}_{1}^{\dagger}}\hat{U}_{1}^{\dagger}}+\overline{\hat{U}_{1}\overline{\hat{U}_{1}}\hat{\rho}\hat{H}_{I}}.
    \label{eq:decoherence_terms_order3}
\end{eqnarray}
Recognizing that $\hat{U}_{1}^{\dagger}=-\hat{U}_{1}$, we can note that equation (\ref{eq:Hamiltonian_third_order_lindblad}) is exactly the same as the equation (\ref{eq:Heff_order3_HI=0}), which leads us to the Hamiltonian defined in (\ref{eq:Heff_order3_harmonic_James}). In this new derivation, the result obtained in (\ref{eq:Heff_order3_harmonic_James}) is verified. Additionally, an expression is provided in order to calculate the decoherence terms, which helps us to provide a more precise description of the dynamic evolution of the system. An important observation given in \cite{James2010}, is that if there is only one frequency in the Hamiltonian, the decoherence terms disappear. It is not difficult to notice in equation (\ref{eq:Heff_order3_harmonic_James}) that for doing the correct dynamical description of a third-order process, it is necessary to consider the decoherence terms due to the existence of more than one frequency in the effective Hamiltonian.
\section{Example}\label{cap:4_secc}
\begin{figure}
\includegraphics[scale=0.6]{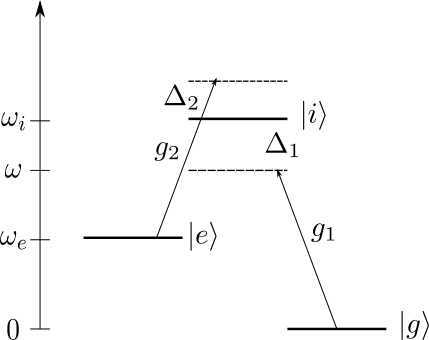}
\caption{\label{fig:level_diagram} Three-level atomic configuration to engineer the effective time-dependent interaction with intensity-dependent coupling.}
\end{figure}

As an illustrative example to test the Hermiticity of the third-order effective time-dependent Hamiltonians given in equations (\ref{eq:Heff_third_order_harmonic_wenjun}) and (\ref{eq:Heff_order3_harmonic_James}), we engineered a time-dependent interaction with an intensity-dependent coupling. To do this, we consider the interaction of an atom with the configuration of levels shown in Fig. (\ref{fig:level_diagram}) and a cavity mode with frequency $\omega$, which drives the transition $\left|g\right\rangle \longleftrightarrow\left|i\right\rangle$ dispersively with the Rabi frequency $g_{1}$ and with a detuning given by $\Delta_{1}=\omega_{i}-\omega$. In the same way, it also drives dispersively the transition $\left|e\right\rangle \longleftrightarrow\left|i\right\rangle$ with the Rabi frecuency $g_{2}$ and a detuning defined by $\Delta_{2}=\omega-\left(\omega_{i}-\omega_{e}\right)$. Additionally, a linear pump with coupling $\lambda$ is applied to the cavity. The Hamiltonian for this system in the interaction picture is ($\hbar=1$)
\begin{equation}
\hat{H}_{I}=g_{1}\hat{\sigma}_{ig}\hat{a}e^{i\Delta_{1}t}+g_{2}\hat{\sigma}_{ie}\hat{a}e^{-i\Delta_{2}t}+\lambda\hat{a}e^{-i\omega t}+\textrm{H.c},
\end{equation}
where $\hat{a}\; (\hat{a}^{\dagger})$ is the operator that describes the annihilation (creation) of a cavity-photon and $\hat{\sigma}_{il}=\left|i\right\rangle \left\langle l\right|$ $ (l=g,e)$ are the operators describing the atomic transitions of the states involved here. In order to apply the effective dynamics up to third order, we identify the following operators: $\hat{h}_{1}=g_{1}\hat{a}^{\dagger}\hat{\sigma}_{gi}$, $\hat{h}_{2}=g_{2}\hat{\sigma}_{ie}\hat{a}$ and $\hat{h}_{3}=\lambda\hat{a}$, with $\omega_{1}=\Delta_{1}$, $\omega_{2}=\Delta_{2}$ and $\omega_{3}=\omega$, note that all frequencies are different. Under the dispersive regime ($\Delta_{1},\Delta_{2}\gg g_{1},g_{2}$ and $\omega\gg\lambda$), the effective dynamic up to third order, according to the method presented in \cite{Shao2017}, is
\begin{eqnarray}
\hat{H}_{W}   (t)=&&\hat{H}_{0}^{\left(2\right)}+\zeta_{eff}\hat{a}\hat{\sigma}_{eg}\hat{a}^{\dagger}\hat{a}e^{i\delta t}+\zeta_{eff}^{\prime}\hat{a}^{\dagger}\hat{a}\hat{\sigma}_{ge}\hat{a}^{\dagger}e^{-i\delta t}\nonumber\\  &&+\xi_{eff}\hat{a}^{\dagger}\hat{a}\hat{\sigma}_{eg}\hat{a}e^{i\delta t}+\xi_{eff}^{\prime}\hat{a}^{\dagger}\hat{\sigma}_{ge}\hat{a}^{\dagger}\hat{a}e^{-i\delta t},
\label{eq:Hw}
\end{eqnarray}
where 
\begin{equation}
    \hat{H}_{0}^{\left(2\right)}=\frac{g_{2}^{2}}{\Delta_{2}}\hat{a}^{\dagger}\hat{a}\hat{\sigma}_{ee}-\frac{g_{1}^{2}}{\Delta_{1}}\hat{a}^{\dagger}\hat{a}\hat{\sigma}_{gg}-\frac{\lambda^{2}}{\omega},
\end{equation}
is the second order contribution to the effective dynamics, $g_{eff}=\lambda g_{1}g_{2}/\Delta_{1}\omega$, $\zeta_{eff}=\omega g_{eff}/\left(\Delta_{1}+\Delta_{2}\right)$, $\zeta_{eff}^{\prime}=\Delta_{1}g_{eff}/\left(\omega-\Delta_{2}\right)$, $\xi_{eff}=-g_{eff}$ and $\xi_{eff}^{\prime}=g_{eff}\left[\Delta_{1}\omega/\Delta_{2}\left(\Delta_{1}+\Delta_{2}\right)+\Delta_{1}\omega/\Delta_{2}\left(\Delta_{2}-\omega\right)\right]$ are the effective atom-field couplings and $\delta=\Delta_{1}+\Delta_{2}-\omega=\omega_{e}-\omega$. Clearly, the effective time-dependent Hamiltonian defined in equation (\ref{eq:Hw}) is not Hermitian, but if $\delta=0$, then the description of the dynamic of the system is still unitary and determined by the effective Hamiltonian
\begin{equation}
\hat{H}=\hat{H}_{0}^{\left(2\right)}+g_{eff}\hat{a}^{\dagger}\hat{a}\left(\hat{\sigma}_{ge}\hat{a}^{\dagger}-\hat{\sigma}_{eg}\hat{a}\right)+\textrm{H.c.}
\label{eq:Hw(0)}
\end{equation}
The expression in equation (\ref{eq:Heff_order3_harmonic_James}), allows us to obtain an effective time-dependent Hamiltonian corresponding to unitary dynamics, given by
\begin{eqnarray}  
\hat{H}_{J}\left(t\right)=&&\hat{H}_{0}^{\left(2\right)}+\Omega_{eff}\left(\hat{a}^{\dagger}\hat{a}\hat{\sigma}_{ge}\hat{a}^{\dagger}e^{i\delta t}+\hat{a}\hat{\sigma}_{eg}\hat{a}^{\dagger}\hat{a}e^{-i\delta t}\right)\nonumber\\
&&+\tilde{\Omega}_{eff}\left(\hat{a}^{\dagger}\hat{\sigma}_{ge}\hat{a}^{\dagger}\hat{a}e^{i\delta t}+\hat{a}^{\dagger}\hat{a}\hat{\sigma}_{eg}\hat{a}e^{-i\delta t}\right),
\label{eq:Hj(t)}
\end{eqnarray}
with $\Omega_{eff}=\left(\zeta_{eff}+\zeta_{eff}^{\prime}\right)/2$ and $\tilde{\Omega}_{eff}=\left(\xi_{eff}+\xi_{eff}^{\prime}\right)/2$. As in the previous case, if $\delta=0$, the Hamiltonian in equation (\ref{eq:Hj(t)}) is equivalent to the one defined in equation (\ref{eq:Hw(0)}). To highlight the usefulness of effective time-dependent dynamics for interaction engineering, we consider a unitary transformation with unitary operator $\hat{U}=\exp\left[-i\hat{H}_{0}^{\left(2\right)}t\right]$ on the Hamiltonian given in equation (\ref{eq:Hj(t)}), such that in this new reference frame it is defined as
\begin{equation}                \hat{H}_{J}^{\prime}\left(t\right)=\sum_{n}\Omega_{n}\left|n+1\right\rangle \left\langle 
  n\right|\hat{\sigma}_{ge}e^{i\phi_{n}t}+\textrm{H.c.},
\label{eq:Hamiltonian_block}
\end{equation}
where
\begin{equation}
 \phi_{n}=\delta-\frac{g_{1}^{2}}{\Delta_{1}}\left(n+1\right)-\frac{g_{2}^{2}}{\Delta_{2}}n,  
\end{equation}
and 
\begin{equation}  \Omega_{n}=\sqrt{n+1}\left[\left(n+1\right)\Omega_{eff}+n\tilde{\Omega}_{eff}\right].  
\end{equation}
The Hamiltonian in equation (\ref{eq:Hamiltonian_block}) is block separable in the subspaces spanned by the states $\left|g,n+1\right\rangle$ and $\left|e,n\right\rangle$ of the atom-field system, and $\phi_{n}$ defines the energy difference between the $\left|g,n+1\right\rangle$ and $\left|e,n\right\rangle$ levels. This energy difference or detuning depends on the number $n$ of photons in the cavity mode. Therefore, for a particular number $m$ of photons in the block defined by subspaces $\left\{ \left|g,m+1\right\rangle ,\left|e,m\right\rangle \right\}$, $\phi=0$ determines the effective resonant frequency of the transition $\left|g,m+1\right\rangle \leftrightarrow\left|e,m\right\rangle$, whereas for the rest of the blocks, those with $n\neq m$ remain in the dispersive regime, that is, $\left|\phi_{m\pm l}-\phi_{m}\right|\gg\Omega_{m\pm l}$ ($l\in\left[1,\infty\right)$ for $\Omega_{m+l}$ and $l\in\left[1,m\right]$ for $\Omega_{m-l}$). It is then possible to remove these highly oscillating terms from the Hamiltonian by using the rotating wave approximation and the atom-field interaction is determined by the selective Hamiltonian with intensity-dependent coupling
\begin{equation}
 \hat{H}_{SI}=\Omega_{m}\left|m+1\right\rangle \left\langle m\right|\hat{\sigma}_{ge}+\textrm{H.c.} 
\end{equation}
The engineering of this Hamiltonian would not be possible if we use the method described in \cite{Shao2017} when the effective resonant frequency of transitions $\left|g,m+1\right\rangle \leftrightarrow\left|e,m\right\rangle$ is due to the counterweight that $\delta$ makes on the terms $\frac{g_{1}^{2}}{\Delta_{1}}\left(n+1\right)$ and $\frac{g_{2}^{2}}{\Delta_{2}}n$.
\section{Summary}
The method for obtaining effective dynamics proposed in \cite{Shao2017}, which is based on the first version of James’ method, clearly requires truncating the Dyson series. However, such truncation does not preserve unitarity. On the other hand, James and Jerke’s time-averaged dynamics method, divides the Hilbert space into parts of low and high frequency. By averaging over time, the high-frequency part of the Hilbert space is eliminated, leading to non-unitary dynamics. This method discards anti-Hermitian terms using the equation (\ref{eq:Heff_hermitian_James}) in order to satisfy the requirement of Hermiticity of the effective Hamiltonian, even if it depends on time. This is not the case for the method used in \cite{Shao2017}, whose Hermiticity is not guaranteed, and then its application is restricted to only a particular case where the effective dynamics of the system under analysis is time-independent. In the latest version of the time-averaged dynamics method presented by Gamel and James \cite{James2010}, the effects generated by suppressing high-frequency terms are considered and then an evolution including decoherence terms is obtained. If these corrections are taken into account, the method provides a more detailed solution, much closer and accurate to the exact solution. Finally, we must remark that in this paper, by using the James and Jerke method \cite{James2007}, an effective third-order Hamiltonian with harmonic time dependence and guaranteed Hermiticity for the effective dynamics with or without time dependence was obtained. For a more accurate description of third-order processes, the decoherence terms of the equation (\ref{eq:decoherence_terms_order3}) must be included because in these circumstances there will necessarily be more than one frequency appearing in the Hamiltonian.


\end{document}